\documentclass[12pt,twoside]{article}

\usepackage{paralist}
\usepackage{enumerate} 

\usepackage{booktabs}
\usepackage{pgfplots}
\usepackage{pgfplotstable}

\newcommand{\createcontingencytable}[4]{ %
\pgfplotstablecreatecol[
    create col/assign/.code={
        \def\rowsum{0}
        \pgfmathtruncatemacro\maxcolindex{\pgfplotstablecols-1}
        \pgfplotsforeachungrouped \col in {1,...,\maxcolindex}{
            \pgfmathsetmacro\rowsum{\rowsum+\thisrowno{\col}}
        }
        \pgfkeyslet{/pgfplots/table/create col/next content}\rowsum
    }
]{#3}{#1}%
\pgfplotstabletranspose[colnames from={#2},input colnames to={#2}]{\intermediatetable}{#1}
\pgfplotstablecreatecol[
    create col/assign/.code={%
        \def\colsum{0}
        \pgfmathtruncatemacro\maxcolindex{\pgfplotstablecols-1}
        \pgfplotsforeachungrouped \col in {1,...,\maxcolindex}{
            \pgfmathsetmacro\colsum{\colsum+\thisrowno{\col}}
        }
        \pgfkeyslet{/pgfplots/table/create col/next content}\colsum
    }
]{#4}\intermediatetable
\pgfplotstabletranspose[colnames from=#2, input colnames to=#2]{\contingencytable}{\intermediatetable}
}
\usepackage{amsmath,amssymb,amsthm,amscd,graphicx,color,accents}

\usepackage[colorlinks=true,linkcolor=blue,urlcolor=blue]{hyperref}
\usepackage{natbib}
\bibpunct{(}{)}{;}{a}{}{,}

\usepackage{caption,lipsum}

\numberwithin{equation}{section}

\theoremstyle{plain}

\def\bP{{\boldsymbol{P}}}
\def\bQ{{\boldsymbol{Q}}}

\def\bLambda{\boldsymbol{\Lambda}}

\def\bLambda{\boldsymbol{\Lambda}}

\def\S{{\mathcal S}}

\begin{document}

\title{\bf\Large{
Long-Term Implications of the Revenue Transfer Methodology in the Affordable Care Act}}

\author{
{Ishan Muzumdar}\thanks{Schreyer Honors College, Pennsylvania State University, University Park, PA 16802.} 
\ {\ and Donald Richards}\thanks{Department of Statistics, Pennsylvania State University, University Park, PA 16802.
}}

\maketitle

\begin{abstract} 
The Affordable Care Act introduced a {\it revenue transfer formula} that requires insurance plans with generally healthier enrollees to pay funds into a revenue transfer pool for to reimburse plans with generally less healthy enrollees.  For a given plan, the issue arises of whether the plan will be a payer into or a receiver from the pool in a chosen future year.  To examine that issue, we analyze data from {\it The Actuary Magazine} on transfer payments for 2014-2015, and we infer strong evidence of a statistical relationship between year-to-year transfer payments.  We also apply to the data a Markov transition model to study annual changes in the payer-receiver statuses of insurance plans.  We estimate that the limiting conditional probability that an insurance plan will pay into the pool, given that the plan had paid into the pool in 2014, is 55.6\%. Further, that limiting probability is attained quickly because the conditional probability that an insurance plan will pay into the pool in 2024, given that the plan had paid into the pool in 2014, is estimated to be 55.7\%.  We also find the revenue transfer system to have the disturbing feature that once a plan enters the ``state'' of paying into the pool then it will stay in that state for an average period of 4.87 years; also, once a plan has received funds from the pool then it will stay in that state for an average period of 3.89 years.

\bigskip 
\noindent 
{{\em Key words and phrases}: Affordable Care Act; Markov transition model; Revenue transfer formula.}

\medskip 
\noindent
{{\em 2010 Mathematics Subject Classification}. Primary: 62P05; Secondary: 60E05.
\smallskip
\noindent
{\em JEL Classification System}.  I13, G22.}
%
\end{abstract}


\section{Introduction}
\label{sec:introduction}

The Affordable Care Act of 2010 introduced a {\it revenue transfer methodology} that was intended to equalize competition among insurance plans.  The revenue transfer methodology was designed to require insurance plans with generally healthier enrollees to pay funds into a revenue transfer pool for the purpose of reimbursing plans with generally less healthy enrollees, thereby dissuading plans from overly favoring healthier enrollees.  We refer to the Patient Protection and Affordable Care Act (2013, pp.~15430--15432) for details of the formula that was designed to calculate revenue transfer amounts.  

For a specific insurance plan, a crucial issue is whether the plan will, in a given future year, be a payer into or a receiver from the revenue transfer pool.  To study that issue, we analyze in this paper the short- and long-term consequences of the revenue transfer methodology.  In particular, we wish to address the long-term feasibility of the methodology, as discussed by Goldman (2016), Wrobel (2016), and others.  

Goldman (2016), in a study of transfer payments to and from the revenue transfer pool in 2014--2015, collected data on a large number of insurance plans in eight states.  Assuming that those plans form a random sample, we perform a chi-square test for statistical independence between year-to-year revenue transfer payments.  We infer from this test overwhelmingly strong evidence of a statistical relationship between the incidence of year-to-year revenue adjustment payments or receipts for ACA plans.

Having found strong evidence of such a statistical relationship, we apply to Goldman's data a Markov transition model to analyze year-to-year changes in the payer-receiver outcomes of insurance plans. The Markov transition model provides the conditional probability that a given insurance plan will be required to pay into the revenue transfer pool in any specific future year given that the plan had paid into the pool in 2014.  We infer that the revenue adjustment methodology, in the long run, and even in the intermediate term, is highly disadvantageous to small ACA insurance plans.  

The paper is organized as follows.  In Section \ref{sec:testingforindependence}, we study the short-term implications of the revenue transfer methodology by performing a test for independence between year-to-year revenue transfer payments.  We determine the $p$-value of the test to be less than $0.00001$ and therefore infer strong evidence of a statistical relationship between an insurance plan's payer-receiver status in 2014 and its payer-receiver status in 2015.  

In Section \ref{sec:thertf}, we apply a Markov transition model to estimate the conditional probability that an insurance plan will be a payer into the revenue transfer pool in a particular future year given that the plan was a payer into the pool in 2015.  We estimate that the limiting conditional probability that an insurance plan will be required to pay into the pool in a very distant future year, given that the plan had paid into the pool in 2014, is 55.6\%.  Further, the ``very distant future'' arrives soon because we also estimate that the conditional probability that an insurance plan will be required to pay into the pool in 2024, given that the plan had paid into the pool in 2014, is 55.7\%.  We also discover that the revenue transfer system has the disturbing feature that once an insurance plan enters the ``state'' of paying into the revenue transfer pool then it will, on average, stay in that state for 4.87 years.  Moreover, once a plan has entered the state of receiving funds from the pool then it will, on average, stay in that state for 3.89 years.  

We provide in Section \ref{sec:implicationsofRTM} a discussion, of the long-term implications of the revenue transfer methodology, in which we summarize in layman's terms the calculations and findings of Section \ref{sec:thertf}; as a consequence of our findings, we find it difficult to escape the conclusion that small ACA plans would be better off either exiting the ACA system or merging with a larger insurance plan.  Finally, we provide in Section \ref{sec:craps} a partly tongue-in-cheek comparison between the revenue transfer methodology and the well-known casino game of craps.  We conclude that, from the viewpoint of capital preservation, long-term ACA ``players'' face far more advantageous odds at the craps tables and hence would be better off taking their money to Las Vegas to play craps on a full time basis.

\section{Testing for independence between year-to-year revenue transfer payments}
\label{sec:testingforindependence}

It is important to health-insurance plans' enrollees and administrations, as with any commercial venture, that there be year-to-year consistency in insurance premium estimates.  Therefore, it would be useful for the purposes of premium calculations that there be a corresponding consistency in the numbers of ACA insurance plans that pay into or receive funds from the revenue transfer pool on a year-to-year basis.  

Goldman (2016) conducted a year-to-year analysis of ACA payments into or from the revenue transfer pool. In particular, Goldman provided data on the numbers of ACA insurance plans that paid into or received funds from the revenue transfer pool in 2014 and 2015. That data were derived from all ACA plans that had greater than 5 million member months and were located in California, Florida, Georgia, Illinois, New York, North Carolina, Pennsylvania, and Texas. After excluding a few very small plans, Goldman (2016) found that there were 113 plans that had revenue adjustment transfers in the individual market in both 2014 and 2015.  By classifying each plan according to its revenue transfer status in 2014 and 2015, Goldman (2016) provided the contingency data in Table \ref{table:nonlin}.  

The data in Table \ref{table:nonlin} show that 87\%, or $55/63$, of the insurance plans that received funds in 2015 from the revenue adjustment pool had also received funds in 2014.  Further, 79\%, or $31/39$, of the insurance plans that paid into the revenue adjustment pool in 2015 had also paid into the pool in 2014.  These observations raise the issue of whether there exists a statistical relationship between the incidence of year-to-year revenue adjustment payments or receipts.

\begin{table}[t]
\caption{Revenue Adjustments in 2014 and 2015 (Goldman, 2016)}
\pgfplotstableread{
    {Payer Status}       {Payer in 2015}    {Receiver in 2015}
    {Payer in 2014}               31                   8           
    {Receiver in 2014}           19                 55          
}\riskadjustmenttransfers
\createcontingencytable{\riskadjustmenttransfers}{Payer Status}{Total}{Total}
\begin{center}
\pgfplotstabletypeset[
  every head row/.style={%
    before row={\toprule 
        & \multicolumn{2}{c}{Revenue Adjustments}\\            \cmidrule{2-3}},
    after row=\midrule},
  every last row/.style={after row=\bottomrule},
  columns/Payer Status/.style={string type},
  columns={Payer Status, Payer in 2015, Receiver in 2015, {Total}},
]\contingencytable
\end{center}
\label{table:nonlin}
\bigskip
\end{table}

In analyzing the data in Table \ref{table:nonlin}, we assume that the data form a random sample from the {\it population of ACA plans that are in those eight states and that have more than 5 million member months}.  

We will apply to the data the Pearson chi-square test for independence between 2014 and 2015 revenue adjustment status.  The null hypothesis for the test is that the incidence of paying or receiving revenue adjustment funds in 2014 is statistically independent of the incidence of paying or receiving revenue adjustment funds in 2015.  The degrees-of-freedom of the chi-square statistic is $1$ and the rejection region for the test is $\chi^2 \ge 3.84$.  

We calculate from the contingency data in Table \ref{table:nonlin} that the observed value of the test statistic is $\chi^2=31.1584$, and the corresponding $p$-value is less than $0.00001$.  Therefore, we reject the null hypothesis.  We infer that, under the current configuration of the revenue transfer methodology, there is strong evidence of a statistical relationship between the incidence of paying or receiving revenue adjustment funds in 2014 and the incidence of paying or receiving revenue adjustment funds in 2015.  

Simply put, we have found strong evidence of a statistical relationship between the incidence of year-to-year revenue adjustment payments or receipts for ACA plans.

\section{A Markov transition model for revenue transfer outcomes}
\label{sec:thertf}

In the sequel, we assume that there is a statistical relationship between the incidence of year-to-year payments or receipts by ACA plans, and now we want to study the long-term implications of that relationship.  

First, we define the ``state'' of an insurance plan in a given year.  We will say that a given health-insurance plan is ``in state $0$'' in a given year if the insurer pays funds in that year into the revenue transfer pool.  The plan will be said to be ``in state 1'' in a given year if the plan receives funds in that year from the revenue transfer pool.  The {\it state space}, $\S = \{0,1\}$, is the collection of possible states.  

We denote by $X_n$, $n = 1,2,\ldots$, a health insurer's state at year $n$.  Thus, for $i \in \S$, an insurance plan is in state $i$ at year $n$ if $X_n = i$.  

Using the data in Table \ref{table:nonlin}, we want to estimate the conditional probability that a given insurer will be a payer into the revenue adjustment pool (i.e., in state $0$) in a particular future year, e.g. 2024, given that the insurer was in state $0$ in 2014.  More generally, we want to estimate the long-term, or limiting, conditional probability that an ACA insurance plan will be required to pay into the revenue adjustment pool in a given year given that it was required to pay into the pool in 2014. 

A standard method for estimating such conditional probabilities is by means of Markov transition models (Ross 2014, Chapter 4).  The Markov model approach supposes that the probabilities of transitions between states $0$ and $1$ exhibit lack-of-memory beyond each current year.  Such a model represents a best-case assumption for receiver-payer transitions in the ACA revenue adjustment process.  Stated alternatively, were it the case that the incidence of having paid in any given year influences the outcome in all subsequent years, such a phenomenon could well be detrimental to those ACA plans which had paid into the pool in 2014, in which case the financial consequences for those plans could be negative.  

In applying a Markov model, we seek to understand how the revenue adjustment system would function over the long term were it to be applied mechanically with all other variables being held fixed.  Thus, we seek to detect the pure influence of the revenue adjustment formula while controlling for the influence of confounding variables, e.g. asymmetry of risk adjuster knowledge and experience, market share, balance sheet fluctuations, and changes in CMS regulations.  As we remark later, we believe that the most important confounding variable is each insurance plan's market share, so we are especially interested in a modeling system that treats insurance plans without explicit regard to size at each revenue adjustment point in time.

In applying the model to Goldman's data, we are making the assumption that the payer-receiver transitions in 2014--2015 generally are representative of the population of all plans and also representative of the pattern of transitions that will ensue over time. These assumptions are standard for Markov model applications.

Under the Markov chain model approach, we suppose that for all $n \ge 1$ and for all $i_0, i_1, \ldots, i_n, i, j \in \S$, 
\begin{equation}
\label{markovmodel}
P(X_{n+1}=j|X_n=i,X_{n-1}=i_{n-1},\dots,X_1=i_1,X_0=0)=P(X_{n+1}=j|X_n=i).
\end{equation}
This is a mathematical formulation of the assumption that the conditional distribution of a future state $X_{n+1}$, given the states $X_n,X_{n-1},X_{n-2},\ldots,X_1,X_0$, is independent of the past states $X_{n-1},X_{n-2},\ldots,X_1,X_0$, and depends only on the present state, $X_n$ (Ross 2014,~p.~184).  
For $i$ and $j$ in $\S$, we define the {\it transition probabilities}, 
$$
P_{ij} = P(X_{n+1}=j|X_n=i).
$$
As examples, $P_{00} = P(X_{n+1} = 0 | X_n = 0)$ is the conditional probability that the insurance plan pays into the pool in year $n+1$ given that the plan was a payer into the pool in year $n$, and $P_{11} = P(X_{n+1} = 1 | X_n = 1)$  is the conditional probability that the insurance plan received funds from the pool in year $n+1$ given that it received funds from the pool in year $n$.  We collect together the transition probabilities in a {\it transition matrix}, 
$$
\bP = \begin{pmatrix}
P_{00} & P_{01} \\ P_{10} & P_{11}
\end{pmatrix}.
$$
The probabilities $P_{ij}$ are also called the {\it one-step transition probabilities} because they represent the probabilities of transitioning between states over a one-year period.  

Based on the data in Table \ref{table:nonlin} we estimate, with accuracy to three decimal places, that 
\begin{equation}
\label{P1matrix}
\begin{array}{ll}
P_{00} = 31/39 = 0.795, \quad & \quad P_{01} = 8/39 \ \, = 0.205 \\
P_{10} = 19/74 = 0.257, \quad & \quad P_{11} = 55/74 = 0.743
\end{array}
\end{equation}

Having defined the one-step transition probabilities, $P_{ij}$, we define the {\it $k$-step transition probability}, $P^{(k)}_{ij}$, to be the probability that a process in state $i$ will transition to state $j$ after $k$ years; that is, 
$$
P^{(k)}_{ij} = P(X_{k+n} = j|X_n = i).
$$
In particular, $P^{(1)}_{ij} = P_{ij}$.  By convention, we also define $P^{(0)}_{ij}$ to equal $1$ if $i=j$ and to equal $0$ if $i \neq j$.  

By the Chapman-Kolmogorov equations (Ross 2014, p. 187), 
\begin{equation}
\label{chapmankolmogorov}
P_{ij}^{(k)} 
= P_{i0}^{(k-r)} P_{0j}^{(r)}  + P_{i1}^{(k-r)} P_{1j}^{(r)},
\end{equation}
for all nonnegative integers $k$ and $r$ such that $r \le k$.  Let 
$$
\bP^{(k)} =\begin{pmatrix}
P_{00}^{(k)} & P_{01}^{(k)} \\ P_{10}^{(k)} & P_{11}^{(k)}
\end{pmatrix}
$$
denote the matrix of $k$-step transition probabilities; then the Chapman-Kolmogorov equations (\ref{chapmankolmogorov}) are equivalent to the matrix multiplication identity, 
\begin{equation}
\label{matrixchapmankolmogorov}
\bP^{(k)} = \bP^{(k-r)} \bP^{(r)} = (\bP^{(1)})^k.
\end{equation}
Starting with the initial one-step transition matrix given in (\ref{P1matrix}), 
$$
\bP^{(1)} = \begin{pmatrix}
0.795 & 0.205 \\
0.257 & 0.743
\end{pmatrix} ,
$$
as estimated from Goldman's data in Table \ref{table:nonlin}, it is straightforward to calculate that the ten-year transition matrix is 
\begin{equation}
\label{P10matrix}
\bP^{(10)} = (\bP^{(1)} )^{10} = 
\begin{pmatrix}
0.557 & 0.443 \\
0.555 & 0.445
\end{pmatrix}.
\end{equation}
%
%
Consequently, we estimate that 55.7\% of all ACA plans in the population that paid into the pool in 2014 will again be payers into the pool in 2024.  Further, we estimate that 44.5\% of all ACA plans in the population that did not pay into the pool in 2014 will again be non-payers in 2024.  The remaining values 44.3\% and 55.5\% are the corresponding complementary probabilities.  


In order to study the long-term consequences of the revenue transfer methodology, we will need an explicit expression for the $k$-step transition matrix $\bP^{(k)}$.   To that end, we derive the eigendecomposition, 
\begin{equation}
\label{eigendecomposition}
\bP^{(1)} = \bQ \bLambda \bQ^{-1},
\end{equation}
where 
$$
\bQ = 
\left(\begin{array}{@{}rr@{}}
0.707 & -0.624 \\
0.707 & 0.781
\end{array}\right)
$$
is the matrix whose columns are the eigenvectors of $\bP^{(1)}$ and 
$$
\bLambda = 
\left({\hskip-4pt}
\begin{array}{ll}
1 & 0 \\
0 &  0.538
\end{array}{\hskip-4pt}\right)
$$
is the matrix whose diagonal entries are the eigenvalues of $\bP^{(1)}$.  Moreover, the $i$th diagonal entry of $\bLambda$ is the eigenvalue of $\bP^{(1)}$ corresponding to the $i$th column of $\bQ$.  By (\ref{matrixchapmankolmogorov}) and (\ref{eigendecomposition}), 
\begin{equation}
\label{Pkmatrix}
\bP^{(k)} = (\bP^{(1)})^k = \bQ \bLambda^k \bQ^{-1} 
= \bQ 
\left(\begin{array}{@{}ll@{}}
1 & 0 \\
0 & (0.538)^k
\end{array}\right) 
\bQ^{-1}.
\end{equation}

We note two important consequences of this formula for $\bP^{(k)}$.  First, it follows by multiplying the matrices on the right-hand side of (\ref{Pkmatrix}) that all entries, $P^{(k)}_{ij}$, of the matrix $\bP^{(k)}$ are positive; therefore, the Markov chain is \textit{irreducible} (Ross 2014, p. 195).  Hence, regardless of whether or not a given ACA plan is required to pay into the revenue adjustment pool in the current year, there is a non-zero probability that it will be required to pay into the revenue adjustment pool in any chosen future year.  

Second, again by multiplying the matrices on the right-hand side of (\ref{Pkmatrix}), it can be verified from (\ref{Pkmatrix}) that for $i = 0$ or $1$, 
\begin{equation}
\label{state-i-recurrent}
\sum_{k=1}^\infty P^{(k)}_{ii} = \infty.
\end{equation}
By Ross (2014, Proposition 4.1, p. 197), it follows from (\ref{state-i-recurrent}) that the states $0$ and $1$ are {\it recurrent}, i.e., each ACA plan will, with probability $1$, transition from any state $i \in \S$ to any state $j \in \S$ at some point in the future.  

As a consequence of general Markov chain theory, or as a direct consequence of (\ref{Pkmatrix}), the $k$-step transition matrix $\bP^{(k)}$ converges as $k \to \infty$ to the limiting transition matrix, 
\begin{equation}
\label{Pinfinitymatrix}
\bP^{(\infty)} = \lim_{k \to \infty} \bP^{(k)} = \bQ 
\left(\begin{array}{@{}ll@{}} 
1 & 0 \\
0 & 0
\end{array}\right)  \bQ^{-1} = 
\begin{pmatrix}
0.556 & 0.444 \\
0.556 & 0.444
\end{pmatrix}.
\end{equation}
In the notation of Ross (2014, p. 204), 
\begin{equation}
\label{pi0andpi1}  
\pi_0 = 0.556,  \quad \hbox{and} \quad \pi_1 = 0.444.
\end{equation}  

In summary, we estimate that the limiting probability that a randomly chosen ACA plan will be required to make a payment into the revenue transfer pool is 55.6\%, and the estimated limiting probability of receiving revenue transfer funds is 44.4\%.  

For each ACA plan, there is also the issue of how long the plan stays in a given state once it enters that state, i.e., the {\it duration time} of the state.  For instance, suppose that an ACA plan was required to pay into the revenue adjustment pool last year; then, we wish to determine how long it can be expected to remain in that state.  Having shown that our Markov chain is irreducible nature, we can use the initial transition matrix (\ref{P1matrix}) and the limiting transition matrix (\ref{pi0andpi1}) to determine the mean number of years that an ACA plan will stay in state 0 or state 1 once it has entered that state.  By proceeding as in Ross (2014, p. 210, Example 4.24), we deduce that once an ACA plan has entered state 0, it will remain in that state for an average period of 
$$
\frac{\pi_0}{\pi_1 P_{10}} = \frac{0.556}{(0.444)(0.257)} = 4.87
$$
years.  Also, we find that once an ACA plan has entered state 1, it will remain in that state for an average period of 
$$
\frac{1-\pi_0}{\pi_1 P_{10}} = \frac{1-0.556}{(0.444)(0.257)} = 3.89
$$
years.  We note that the difference between the two mean duration times is one year, approximately.

\section{The long-term implications}
\label{sec:implicationsofRTM}

Starting from the one-step transition matrix $\bP^{(1)}$ derived from the data in Table 1, we obtained the ten-year transition matrix $\bP^{(10)}$ in (\ref{P10matrix}).  We obtain an estimate of 55.7\% for the probability that a randomly chosen ACA plan that paid into the revenue adjustment pool in 2014 will be required to pay into the pool in 2024.  Further, we obtain an estimate of 44.5\% for the probability that a randomly chosen ACA plan that did not pay into the pool in 2014 will not be required to pay into the pool in 2024.  

As noted by the American Academy of Actuaries (2016), Li and Richards (2017), and others, there were clear disparities in size between those ACA plans that paid into the pool in 2014 as compared to those plans that did not.  Specifically, plans that were required to pay into the pool in 2014 often were smaller both in membership size and capitalization, whereas plans that were not required to pay into the pool generally had larger memberships and greater capitalization.  Given the substantial difference between the corresponding estimated ten-year transition probabilities of 55.7\% and 44.5\%, it appears that smaller ACA plans may be significantly disadvantaged under the revenue transfer methodology over ten year periods.  

We showed further that the Markov chain is irreducible.  A crucial consequence of the property of irreducibility is that any given ACA plan may be required to pay into the revenue adjustment pool in any chosen future year regardless of whether the plan is required to pay into the pool in the current year.  Moreover, the property that the Markov chain is recurrent implies that each ACA plan will, almost surely, transition at some point in the future from its current state to the opposite state.  Such a system may seem judicious to a health insurance regulator; however, the system has the drawback that the likelihood of a given plan having to pay into the pool in a given year is markedly higher than the probability that it will receive funds from the pool in that year.  Therefore, over the long term, the plan is at higher risk of financial stress simply because of the revenue transfer methodology.  

Turning to the long-term transition matrix, we estimated that the limiting probability that a randomly chosen ACA plan will be required to make a payment into the revenue transfer pool is 55.6\%, and the estimated limiting probability of receiving revenue transfer funds is 44.4\%. These long-term estimated probabilities imply that a substantially greater percentage of health care plans will be required to pay into the revenue transfer pool than to receive funds from it.

We also remark that the limiting (or perpetual) state arrives quickly.  By applying (\ref{Pkmatrix}), we obtain the eight-year transition matrix, 
$$
\bP^{(8)} = 
\begin{pmatrix}
0.559 & 0.441 \\
0.552 & 0.448
\end{pmatrix}.
$$
which is startlingly close to $\bP^{(\infty)}$, the limiting transition matrix given in (\ref{Pinfinitymatrix}).  So it appears that eight years is approximately the length of time it would take the system to attain the limiting state.  

Exacerbating these pitfalls of the revenue transfer methodology are the duration times for the payer and non-payer states.  As shown at the end of Section \ref{sec:thertf}, there is a difference of approximately one year in the mean duration time for plans paying into the pool to remain in the ``paying'' state as compared to plans receiving funds from the pool to remain in the ``receiving'' state.  Therefore, the Markov model analysis reveals that the revenue transfer methodology may demonstrate features of a system in which the ``rich'' are likely to become richer and the ``poor'' are likely to become poorer.  


As shown by Li and Richards (2017) smaller ACA insurance plans, simply because they are small, face serious obstacles from the revenue adjustment system.  In addition, small plans face the usual substantial difficulties that pertain to any small insurance operation, e.g., lesser capital that make them less able to withstand balance sheet downturns and greater exposure to executive turnover, changes in CMS regulations, changes in the numbers of participating plans, and population changes; see also Perlman and Liner (2016) and Siegel and Liner (2015) for extensive discussions of the actuarial disadvantages of the ACA for small insurance plans.  Going beyond those disadvantages, an implication of our Markov model analysis is that once an ACA plan falls into the category of paying into the ACA revenue adjustment pool, the plan is likely to remain in the pool for a longer period than it is likely to remain out of the pool. Hence, we infer that, even under the best-case Markov model scenario, smaller ACA plans are likely to face continued financial difficulties.  

In conclusion, our advice to small insurers is: Either exit the ACA system or merge quickly with a larger insurer.

\section{A comparison with the game of craps}
\label{sec:craps}

The game of craps is a gambling game that is popular in casinos across the U.S.  The game is instructive for teaching elementary probability theory to students (Blake 1979, pp. 87-88).  As the probability of a gambler winning on each play of craps is $244/495$, or 49.3\%, the game is also known to be the casino game that is closest to fair.  Were we to regard the ACA revenue transfer system as a ``game'' in which ``winning'' corresponds to receiving funds from the revenue adjustment pool, then we have estimated that the probability of ``winning'' on each annual attempt is 44.4\%.  It is difficult to escape the conclusion that long-term ACA ``players'' would be better off financially by taking their money to Las Vegas to play craps on a full-time basis.  

\bigskip
\bigskip
\smallskip

\noindent
{\bf Acknowledgment}: 
The authors thank  the reviewers and the editor for their constructive  comments.

\bigskip

\noindent
{\bf Disclaimer}:
The authors wish to report that there are no funding sources or other affiliations that may represent a conflict of interest in the results presented in this paper.  The views expressed here are those of the authors and are not necessarily those of their employers.  This paper has not been published elsewhere, and it has not been submitted simultaneously for publication elsewhere.

\bigskip

\noindent
{\bf Correspondence}: 
Donald Richards, Department of Statistics, 326 Thomas Building, Penn State University, University Park, PA 16802; \href{mailto:richards@stat.psu.edu}{richards@stat.psu.edu}, Tel. (814) 865-3993, Fax. (814) 863-7114.

\vspace{5mm}


\newenvironment{reflist}{\begin{list}{}{\itemsep 0mm \parsep 0.2mm
\listparindent -5mm \leftmargin 5mm} \item \ }{\end{list}}
%




\end{document}